\documentclass[aps,prl,floats,twocolumn,epsf]{revtex4}

\usepackage{graphicx}
\usepackage{dcolumn}
\usepackage{bm}
\begin{document}

\title{
Detection of Striped Superconductors Using Magnetic Field Modulated Josephson Effect
}

\author{Kun Yang}

\affiliation{Department of Physics, Florida State University, Tallahassee, Florida 32306, USA
}

\date{\today}

\begin{abstract}
In this paper I show how to use magnetic field modulated Josephson effect to probe the order parameter spatial structure of striped superconductors, proposed by Berg and co-workers.  A possible example of this is $La_{2-x}Ba_xCuO_4$, where it was proposed that the superconducting order parameter oscillates within each Cu-O plane in a manner similar to the FFLO state, and the oscillation pattern is staggered from one layer to another.\\

\end{abstract}

\pacs{Keywords: FFLO, Josephson Effect}

\maketitle

In an interesting Letter\cite{berg},
Berg {\em et al.} suggested that a novel form of superconducting state is realized in La$_{2-x}$Ba$_x$CuO$_4$ with $x$ close to $1/8$. This suggestion was based on experiments\cite{li} on this compound which found predominantly two-dimensional (2D) characters of the superconducting state, with extremely weak interplane coupling. Later this specific form of superconducting state was termed striped superconductors\cite{berg08}. The purpose of this paper is to point out that the suggested form\cite{berg} of the superconducting order parameter can be detected directly using magnetic field modulated Josephson effect.

The importance of charge and spin ordering notwithstanding, the most distinct feature of the state proposed in Ref. \onlinecite{berg} is that the superconducting order parameter is oscillatory with zero mean within each CuO plane; see Fig. 1 for an illustration. This is very similar to what happens in the Fulde-Ferrell-Larkin-Ovchinnikov (FFLO) superconducting states\cite{ff,lo}. It was pointed out some years ago that the order parameter structure of the FFLO states can be probed directly using magnetic field modulated Josephson effect\cite{kun00}. We point out that the same ideas can be applied to striped superconductors, and they can be implemented in two different but related ways.

\begin{figure}
\includegraphics[width=0.45\textwidth]{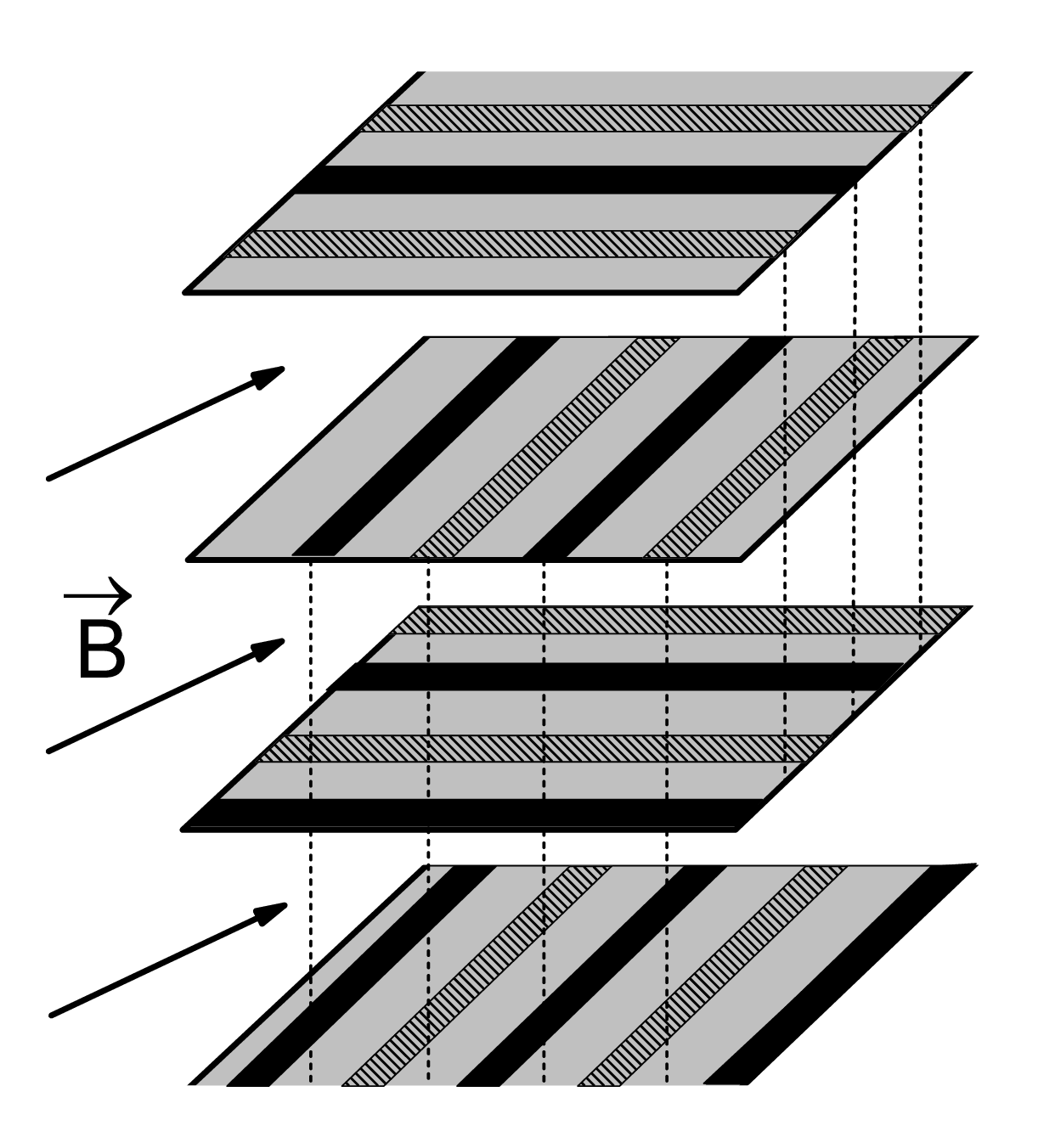}
\caption{Illustration of possible striped superconducting state in $La_{2-x}Ba_xCuO_4$, and possible enhancement of Josephson tunneling between neighboring layer by an external magnetic field. The stripe configuration is rotated by 90 degrees between neighboring CuO layers. The superconducting order parameter oscillates along the direction perpendicular to the stripes. A magnetic field along the direction 45 degrees from the stripes (partially) compensates for the momentum difference between superconducting orders in neighboring layers, thus enhancing Josephson tunneling.}
\end{figure}

1. Consider a Josepheson junction between a striped superconductor and an ordinary superconductor with spatially uniform order parameter (but with the same internal symmetry, say d-wave), with the junction parallel to the CuO plane (such that tunneling is along the $\hat{c}$ direction). See Fig. 2 for an illustration. As pointed out in Ref. \onlinecite{kun00}, due to the mismatch of order parameter momenta in the two superconductors, Josephson effect is suppressed. However Josephson effect can be {\em restored} by applying a magnetic field parallel to the junctions, such that the momenta mismatch is canceled by the phase modulation of Josephson coupling by the magnetic field (see also Fig. 1a of Ref. \onlinecite{kun00}). The condition for the restoration for the magnetic field is\cite{kun00}
\begin{equation}
{\bf B}_0={\hbar c\over 2ed}\hat{c}\times {\bf k},
\end{equation}
where $d$ is the thickness of the junction, and ${\bf k}$ is the (dominant) momentum of the Cooper pair of the striped superconductor. For the state proposed in Ref. \onlinecite{berg}, $|{\bf k}|=\pi/(4a)$ where $a$ is the lattice constant in the CuO plane. Assuming Josephson coupling is dominated by the top CuO layer, the direction of ${\bf B}$ should be parallel to the stripe direction in that layer, which is along either the $\hat{a}$ or $\hat{b}$ direction. To have the magnitude of ${\bf B}_0$ within experimentally accessible range (say $10T$ or below), we need the junction thickness $d \gtrsim 100nm$. In reality the junction thickness will most likely be (significantly) smaller than this, implying the desired magnitude $B_0$ is out of reach. In this case there will still be a field-dependent Josephson critical current, as long as the junction size is finite. Assuming a rectangular-shaped junction with cross section area $A$ covering exactly an integer number of periods of superconducting order parameter, we expect the critical current to be the sum of two {\em shifted} Fraunhofer patterns:
\begin{eqnarray}
&I_c&\propto |\sin[(\pi A(B+B_0)/(\pi\Phi_0)]/[(\pi A(B+B_0)/(\pi\Phi_0)]\nonumber\\
&+&\sin[(\pi A(B-B_0)/(\pi\Phi_0)]/[(\pi A(B-B_0)/(\pi\Phi_0)]|,
\end{eqnarray}
where $\Phi_0$ is (Cooper pair) flux quantum. The two terms correspond to contributions from the two dominant Fourier components of the order parameter; we have neglected possible contributions from (subleading) higher harmonics. In principle one can extract $B_0$ and therefore ${\bf k}$ from the $B$ dependence of $I_c$.

\begin{figure}
\includegraphics[width=0.45\textwidth]{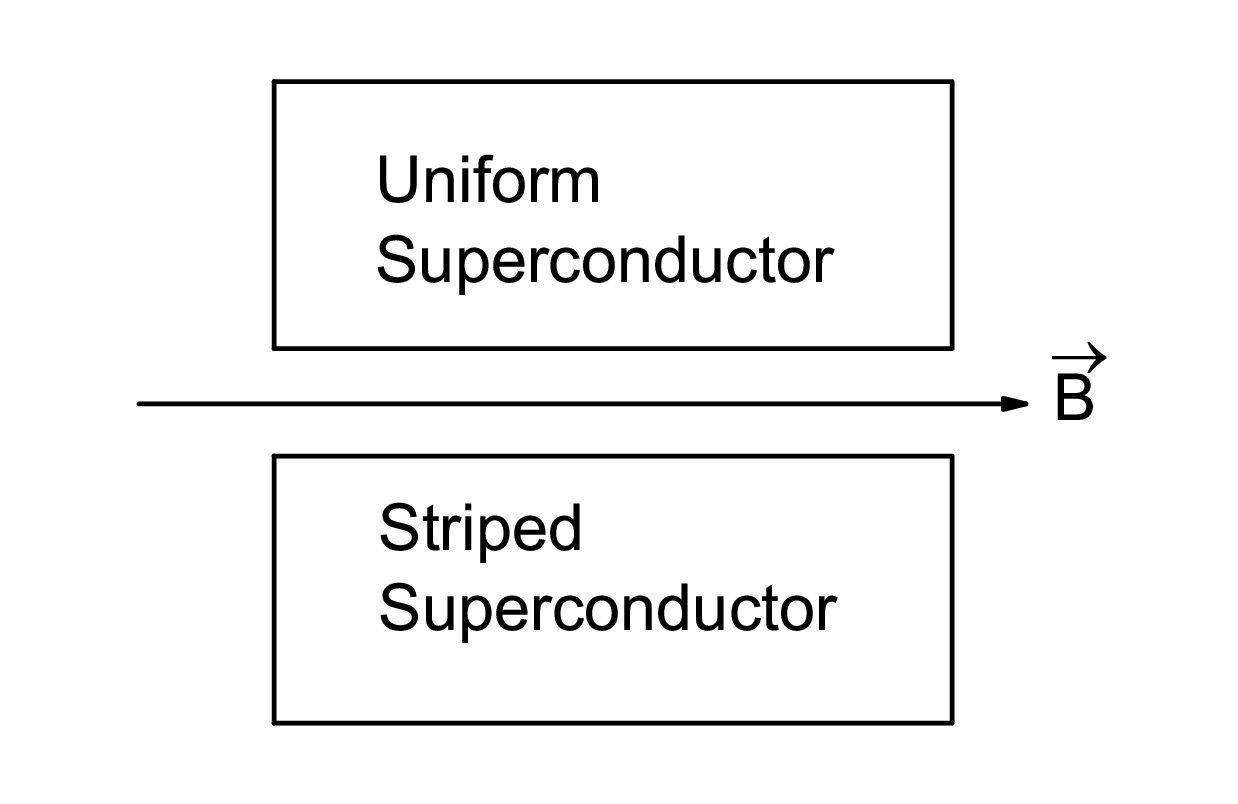}
\caption{A Josephson junction between a striped superconductor and a uniform superconductor.}
\end{figure}

2. Due to the alternation of the stripe direction in neighboring CuO planes, Josephson coupling between the layer is suppressed for a similar reason\cite{berg}. This is argued to be the source of the suppression of interplane coupling in the system of Ref. \onlinecite{li}. Using the same reasoning as in point 1, one can in principle restore the interplane coupling, and dramatically increase bulk superconducting  temperature, by applying a properly chosen magnetic field; see Fig. 1. The equation (1) still applies, except now $d$ is interlayer spacing, and ${\bf k}$ is momentum {\em difference} between the superconducting order parameters in neighboring CuO layers. This implies ${\bf B}_0$ should be applied along the {\em diagonal} (or 110) direction, and the magnitude for restoration of interplane Josephson coupling is of order $1000T$. While this field is too high to reach (and the system would have been destroyed by the field anyway), superconductivity may get enhanced by much lower fields when applied in the correct (or 110) direction, and this might show up in transport or other properties.

We note in closing that due to their similarity mentioned earlier, other methods proposed to detect FFLO-type superconductors might also be used to probe striped superconductors\cite{ym}.

The author thanks Steve Kivelson and Shou-Cheng Zhang for useful discussions and encouragements. This work was supported by National Science Foundation grant No. DMR-0704133, and DMR-1004545.

\end{document}